\documentclass[fleqn,usenatbib]{mnras}

\usepackage{newtxtext,newtxmath}

\usepackage[T1]{fontenc}
\usepackage{ae,aecompl}

\newcommand{\MSun}{\mbox{${\rm M}_\odot$}}
\newcommand{\Msun}{\mbox{${\rm M}_\odot$}}

\usepackage{graphicx}	
\usepackage{amsmath}	
\usepackage{amssymb}	

\def\aplt{\ {\raise-.5ex\hbox{$\buildrel<\over\sim$}}\ }
\def\apgt{\ {\raise-.5ex\hbox{$\buildrel>\over\sim$}}\ }

\title[wind ripples]{The Signature of a Windy Radio Supernova Progenitor in a Binary System}

\author[Almog Yalinewich \& Simon Portegies Zwart]{
Almog Yalinewich,$^{1}$\thanks{E-mail: almog.yalin@gmail.com}
Simon Portegies Zwart$^{2}$
\\
$^{1}$Canadian Institute for Theoretical Astrophysics, 60 St. George St., Toronto, ON M5S 3H8, Canada\\
$^{2}$Leiden Observatory, Leiden University, Niels Bohrweg 2, NL
-2333 CA Leiden, the Netherlands\\
}

\date{Accepted XXX. Received YYY; in original form ZZZ}

\pubyear{2018}

\begin{document}
\label{firstpage}
\pagerange{\pageref{firstpage}--\pageref{lastpage}}
\maketitle

\begin{abstract}
Type II supernova progenitors are expected to emit copious amounts of mass in a dense stellar wind prior to the explosion. When the progenitor is a member of a binary, the orbital motion modulates the density of this wind. When the progenitor explodes, the high-velocity ejecta collides with the modulated wind, which in turn produces a modulated radio signal. In this work we derive general analytic relations between the parameters of the radio signal modulations and binary parameter in the limit of large member mass ratio. We use these relations to infer the semi major axis of SN1979c and a lower bound for the mass of the companion. We further constrain the analytic estimates by numerical simulations using the AMUSE framework. In these calculations we simulate the progenitor binary system including the wind and the gravitational effect of a companion star. 
The simulation output is compared to the observed radio signal in supernova SN1979C.
We find that it must have been a binary with an orbital period of about 2000\,yr. If the exploding star evolved from  a $\sim 18$\,\MSun\, zero-age main-sequence at solar metalicity, we derive a companion mass of $5$ to $12$\,\MSun\, in an orbit with an eccentricity lower than about 0.8.
\end{abstract}

\begin{keywords}
keyword1 -- keyword2 -- keyword3
\end{keywords}



\section{Introduction}
One of the open questions in stellar astrophysics concerns the identity and properties of supernova progenitors \citep{Smartt2009ProgenitorsSupernovae, Maeda2016ProgenitorsSupernovae}. Despite the large number of supernovae detected, little can be inferred about the progenitors because peak luminosities occur after the progenitor has exploded. In some rare cases the progenitor of a nearby supernova can be identified from archival data \citep{VanDyk2017TheProgenitors}. Based on population studies, many supernova progenitors have a companion which can alter the progenitor's evolutionary track \citep{PortegiesZwart1999ThePulsars,Sana2012BinaryStars}.

In this work we describe a method to infer information not just about the progenitor, but also about its binary companion and its orbit. Some supernova progenitors emit copious amount of stellar wind prior to the explosion \citep{Heuvel2013AreClusters,Moriya2014Mass-lossExplosion}. The motion of the progenitor around the centre of mass modulates the wind, increasing the density in the forward direction (relative to the velocity of the progenitor) and reducing it in the backward direction \citep[see also][]{Ishii1993NumericalGas,Saladino2018GoneSystems}. When the progenitor explodes, the ejecta collide with the ripples in the wind and launch a shock wave. A fraction of the thermal energy in the shock is converted into magnetic fields and supra thermal electrons. These relativistic electrons gyrate in the magnetic fields and emit synchrotron radiation in radio \citep{Chevalier1998SynchrotronSupernovae}. As a result of the modulation in the wind, the flux of the radio emission will also fluctuate. In this paper we relate the fluctuation in radio to the density modulation in the wind, and then to the properties of the progenitor binary system.

Fluctuations in the radio have been measured for SN1979c \citep{Weiler1991The1979C}. This supernova has been followed up for about ten years, and in that period the radio flux fluctuated with a period of about four years, and a relative amplitude of about 0.1.  Based on later X-ray observations it was suggested that the remnant of the supernova is a 5-10 $M_{\odot}$ black hole \citep{Patnaude2009Evidence1979C}, and the mass of the progenitor is estimated at $18\pm 3 M_{\odot}$. The mass loss rate close to the explosion time was estimated at $\sim 1.2 \cdot 10^{-4} M_{\odot}/\rm y$ \citep{Weiler1991The1979C}. With such conditions, SN1979c is a good test case to demonstrate how binary parameters can be inferred from variations in the observed radio signal.

This paper is organised as follows, in the next section we derive the analytic formalism which we apply in \S\,\ref{Sect:SN1979C} to the observed supernova SN1979C. In \S\,\ref{Sect:simulations} this formalism is tested by means of simulations and we conclude in \S\,\ref{Sect:discussion}.

\section{Analytic Estimates}

Let us consider two stars with masses $M \gg m$ in a binary with orbital period $P_{orb}$. The stars are assumed to move on circular orbits ($e=0$) with a semi major axis $a$ and observed edge on. The period is given by

\begin{equation}
    P_{orb} = 2 \pi \sqrt{\frac{a^3}{G \left(M+m\right)}} \, .
\end{equation}

The radius of the orbit of the massive star is $am/M$, and therefore its orbital velocity is

\begin{equation}
    v_{M} = \frac{m}{M} \sqrt{\frac{G \left(M+m \right)}{a}} \, .
\end{equation}

We also assume that the more massive member emits a wind at velocity $v_w$. The relative amplitude of the density of the wind due to the motion of the massive companion scales with the ratio of orbital to wind velocity

\begin{equation}
    \frac{\delta \rho}{\rho} \approx \frac{v_M}{v_w} \, .
\end{equation}

Now let us consider what happens when companion $M$ explodes. We assume that as a result of the explosion, the ejecta expand as a spherical shell moving at $v_e \gg v_w$, and collides with previously emitted wind. As a result, a shock wave emerges and travels into the wind with a velocity comparable to $v_e$. This shock emits in the radio, and the strength of the emission oscillates with the upstream density. Since the ejecta travel faster then the wind, then the observed period of the oscillations will be smaller than the orbital period
\begin{equation}
    P_{obs} \approx P_{orb} \frac{v_w}{v_e} \, . \label{eq:observed_period}
\end{equation}
If we assume that the relative fluctuations in flux are comparable to the fluctuations in density, we get
\begin{equation}
    \frac{\delta f}{f} \approx \frac{v_M}{v_w} \, . \label{eq:flux_ratio}
\end{equation}
In the limit of a small secondary $M\gg m$, we can invert equations \ref{eq:observed_period} and \ref{eq:flux_ratio} to get the secondary mass and separation
\begin{equation}
    a \approx \left(\frac{P_{obs}}{2 \pi} \frac{v_e}{v_w}\right)^{2/3} \left(G M\right)^{1/3}. \label{eq:inferred_a}
\end{equation}

\begin{equation}
    m \approx M \frac{\delta f}{f} \left(\frac{P_{obs} v_w^2 v_e}{2 \pi G M}\right)^{1/3} \, . \label{eq:inferred_M}
\end{equation}

\subsection{Synchrotron Emission}

In this section we calculate the prefactor of equation \ref{eq:flux_ratio}. To do so, we assume that the emission is due to synchrotron emission \citep{Rybicki1979RadiativeAstrophysics}, where the generation of magnetic field and acceleration of particles to relativistic energies consumes a certain fraction of the downstream thermal energy \citep{Chevalier1998SynchrotronSupernovae}. The net energy flux per particle is given by
\begin{equation}
    L_s \approx B^2 r_e^2 c \gamma^2,
\end{equation}
where $B$ is the magnetic field, $r_e$ is the classical electron radius, $c$ is the speed of light and $\gamma$ is the Lorentz factor of the electron. Most of the energy is emitted at the synchrotron frequency
\begin{equation}
    \omega_s \approx \frac{q B}{m_e c} \gamma^2
\end{equation}
where $q$ is the elecron charge and $m_e$ is its mass. The magnetic energy is assumed to be a fixed fraction of the thermal energy, so $B^2 \propto \rho v_e^2$. The net luminosity per unit frequency therefore changes as
\begin{equation}
    L \propto \rho L/\omega_s \propto \rho^{3/2}.
\end{equation}
Therefore, the relation between the relative changes in observed flux to relative changes in density is
\begin{equation}
    \frac{\delta f}{f} \approx \frac{3}{2} \frac{\delta \rho}{\rho} \, .
\end{equation}

\section{Application to SN1979c}\label{Sect:SN1979C}

If we assume zero inclination and zero eccentricity, and a large ratio between the members of the binary, we can use equations \ref{eq:inferred_a} and \ref{eq:inferred_M} to estimate the semi major axis of the binary system and the mass of the star accompanying the supernova progenitor.

\begin{equation}
    a \approx 430 \tilde{P}_{obs}^{2/3} \tilde{v}_e^{-2/3} \tilde{v}_w^{2/3} \tilde{M}^{1/3} \rm au, 
\end{equation}

\begin{equation}
    m \approx 6 \tilde{F} \tilde{M}^{2/3} \tilde{P}_{obs}^{1/3} \tilde{v}_w^{2/3} \tilde{v}_e^{1/3} \,  M_{\odot},
\end{equation}
where $\tilde{P}_{obs} = P_{obs}/4 \rm \, year$, $\tilde{v}_e \approx v_e/10^4 \rm \, km/s$, $\tilde{v}_w \approx v_w/20 \rm \, km/s$, $\tilde{M} = M/20 M_{\odot}$ and $\tilde{F} = 10\cdot \delta f/f$.

We note that the mass obtained here is only a lower limit on the actual companion mass. Inclination can reduce the apparent amplitude of the radio oscillation by a factor $\propto \cos i$. Since our expression for the mass scales with the apparent amplitude of the radio oscillation, the real mass of the companion star may be larger than the estimated mass by a similar factor. However, as we demonstrate in the next section this factor does not propagate linearly into the mass estimate.

\section{Simulations}\label{Sect:simulations}

Having derived first order estimates for the binary parameters that could lead to the observed ripples in the radio flux for SN1979C, we now perform a series of simulations in which we verify some of the earlier analytical estimates and constrain the parameters of the progenitor system further. For this purpose we simulate the binary system prior to the supernova, assuming that the supernova blast-wave interacting with the progenitor's wind is responsible for the observed fluctuations.

We simulated the wind evolution of the binary system over the 14,000 years prior to the supernova. The simulations ware performed using the Astrophysical Multipurpose Software Environment ({\em AMUSE} for short) \citep{PortegiesZwart2012Multi-physicsInterface,PortegiesZwart2011AMUSE:Environment,PortegiesZwart2018AMUSE:Environment}. {\em AMUSE} is a component library for multiscale and multiphysics simulations. We set-up our simulation using a binary system with an 18\,\MSun\, Solar metalicity primary at zero age. The primary star was evolved for 
9.6\,Myr using the MESA \cite{Paxton2010ModulesMESA,Paxton2010MESA:Astrophysics} stellar evolution code at which time the star exhausted the central fuel in its 5.27\,\Msun\, core. At this time the star has a luminosity of $\sim 1.1 \cdot 10^{4}$\,L$_\odot$ a surface temperature of about 3500\,K and a mass loss rate in its wind of $\sim 1.0 \cdot 10^{-4}$\,\MSun/yr.

We ran simulations with companion masses  1, 3, 6, 9 12 or 15\,\MSun. The companion star was evolved to the same age as the primary before putting it a binary system. We performed several other calculations in which we varied the orbital period ($P= 500$ yr, 1500 yr and 2000 yr) and eccentricity ($e=0.15$, 0.45, 0.60, 0.75 and $e=0.8$), but the results turn out to be rather insensitive to changes in the eccentricity or the companion mass.

We run the wind module in {\em AMUSE} which was designed to simulate the slow accelerating wind of an AGB star \cite[see van der Helm et al, in preparation, ][]{Saladino2018GoneSystems} with the {\em Fi} smoothed particles hydrodynamics solver \citep{Gerritsen1997StarFormation.,Pelupessy2004PeriodicGalaxies}.
The integration of the equations of motion for the stars is realised using the {\tt Huayno} symplectic $N$-body integrator \citep{Pelupessy2012N-bodySplitting}.
the two codes are coupled with a bridge \citep{PortegiesZwart2018AMUSE:Environment} using a time step of 1 year.
We started the calculation with 50\,SPH particles in a sphere around the primary star, and produce new particles at runtime with a mass of $10^{-5}$\,\MSun. We performed additional simulations with $10^{-6}$\,\MSun\, and $10^{-7}$\,\MSun\, per SPH particle, but the results are rather insentive to the adopted resolution. In the highest resolution, however, we found interesting substructure in the wind, probably due to turbulent motion, but for our comparison with the observations this is irrelevant. The particles are released in a homogeneous sphere with a radius of 100\,au around the primary star, and ejected using to the {\em accelerated} wind model in {\em AMUSE} with an adopted terminal velocity of 10\,km/s (which is consistend with the estimated velocity from the observations \citep{Weiler1991The1979C}).
Due to internal pressure and the local heating by the primary star, the actual terminal velocity in the wind is sometimes hard to constrain, but will result naturally from the hydrodynamical simulation. As a consequence, the terminal velocity of the wind in the simulation is closer to 23\,km/s, which is consistent with known wind velocities of comparable progenitors \citep[see for example,][]{Beasor2018TheRates}.

\begin{figure}
\centering
\includegraphics[width=\columnwidth]{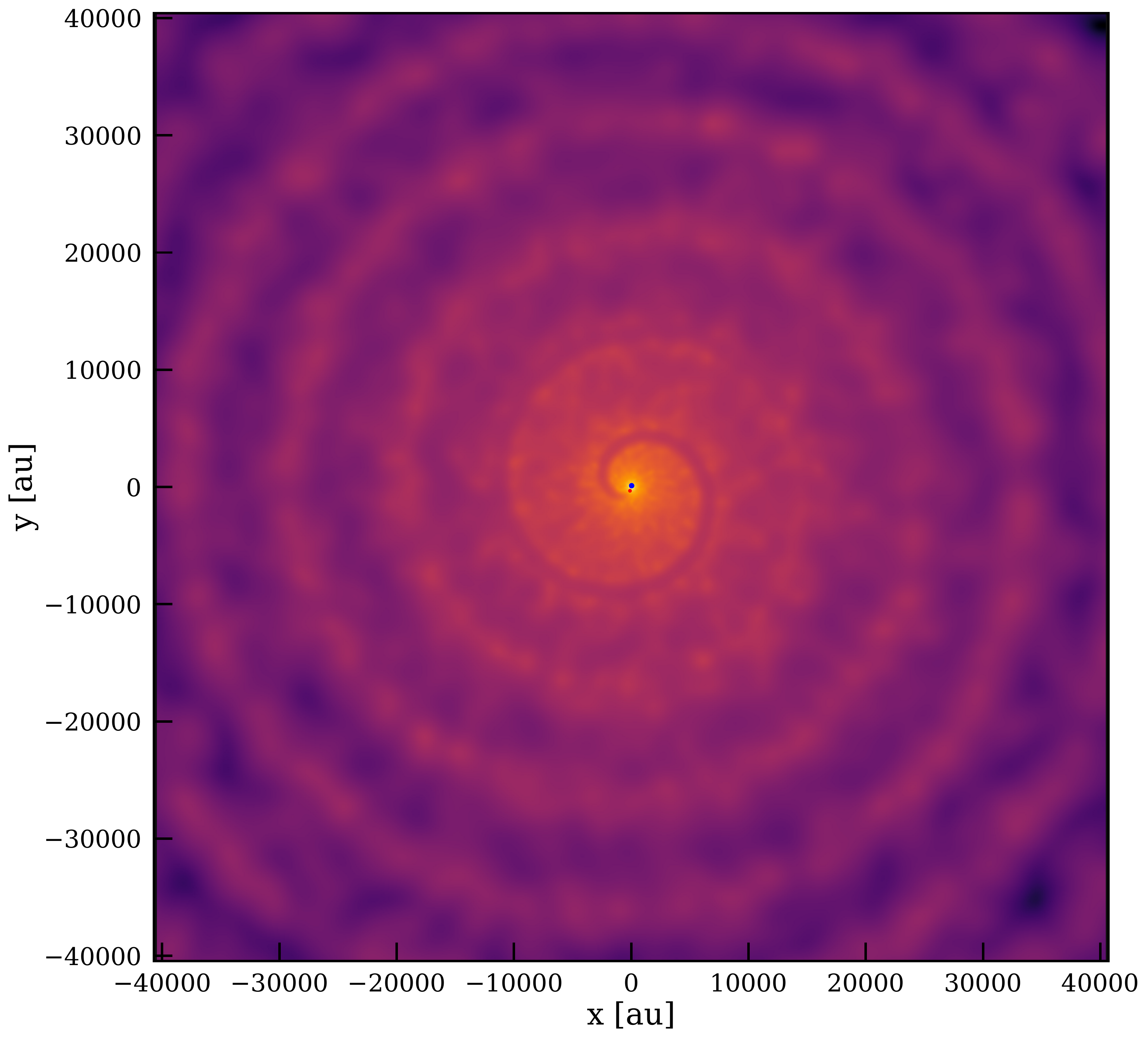}
~\includegraphics[width=\columnwidth]{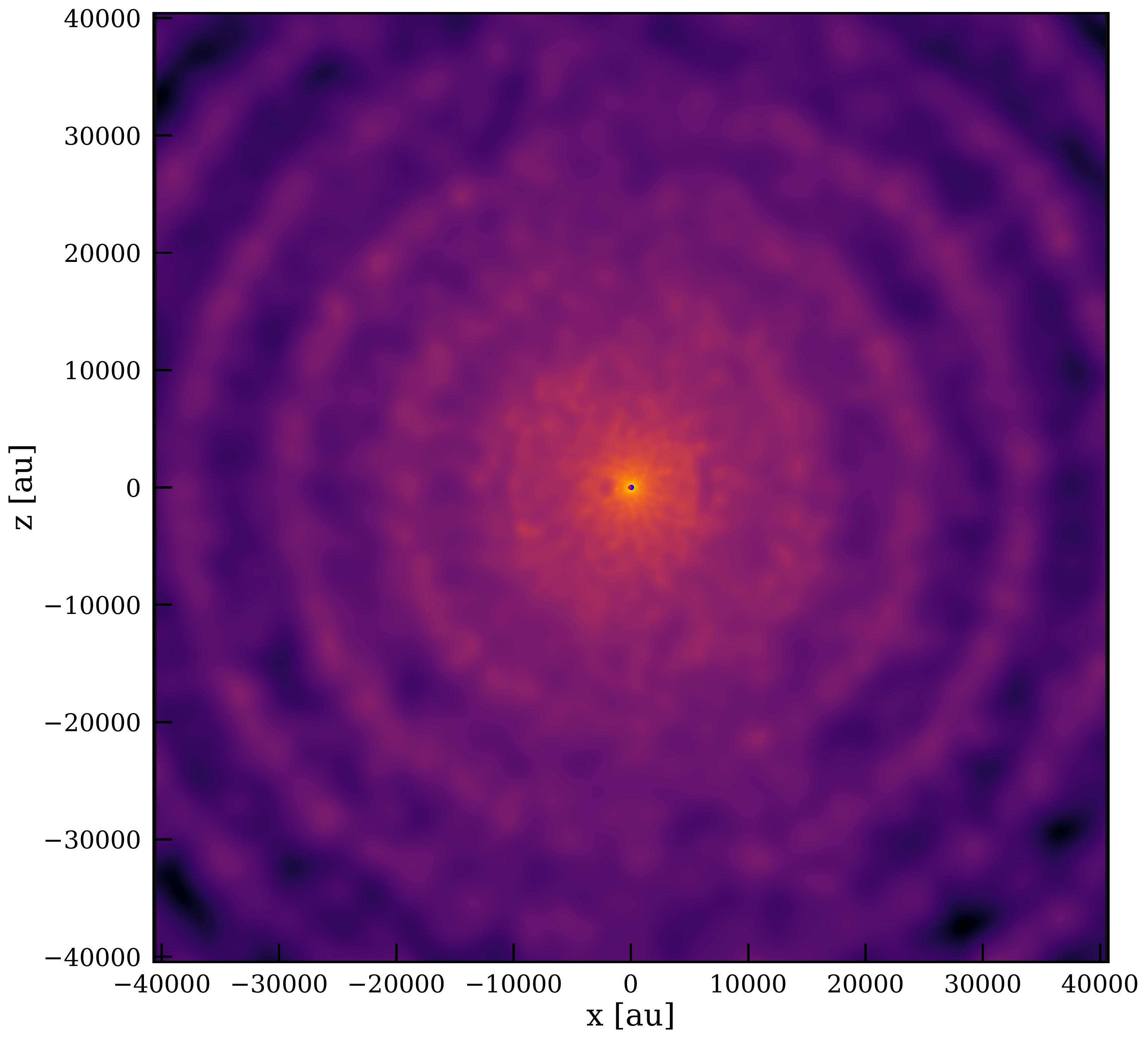}
\caption{Face on (top) and edge on (bottom) view of the gas density distribution 14,000 years after the beginning of wind injection. Both images are centred on the supernova progenitor. 
The primary (blue bullet) is a 18\,\MSun\ supernova progentor and the companion is a 6\,\Msun\ (red bullet). The orbit is circular with a period of 2000 years. The mass per SPH particle is $10^{-5}$\,\MSun. }
\label{fig:SN1979C_M15MSunP2000yr_e0_m-6M}
\end{figure}

We run the models for 14,000 years producing snapshots every 100 years. In fig.\,\ref{fig:SN1979C_M15MSunP2000yr_e0_m-6M} we show the ripples caused by the stellar wind in the final shapshot using a 6\,\Msun\, donor in a 2000yr circular orbit.
The relatively high mass, compared to our earlier estimate was adopted because it turned out to produce an excellent comparison with the observed ripples, but similarly well matching results could be obtained also for a lower mass companion star.

\begin{figure}
\centering
\includegraphics[width=\columnwidth]{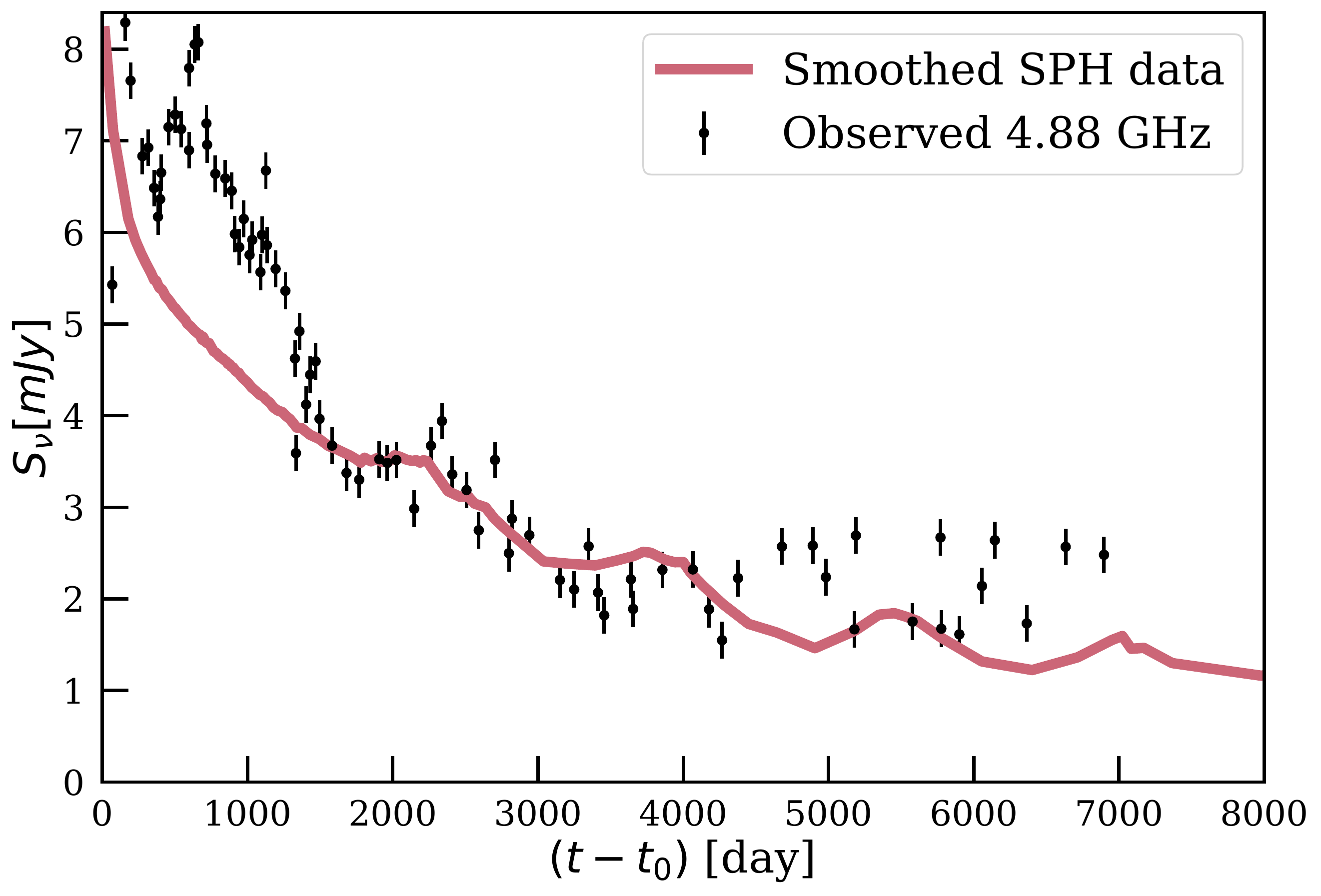}
\caption{The 4.88GHz radio flux as a function of time observed for SN 1979C (black bullets) taken directly from figure 2 of \citep{Weiler1991The1979C}.  The  simulated curve is calculated using a 6\,\MSun\, companion star in a circular orbit of 2000\,yr. We shifted the curves to match them up with the observed radio flux (by $\log \rho = 22.54$ in units of $g/cm^3$). 
The first 5000 days after the onset of the supernova is corrected with a power-law of index 1.6 to match up the baseline of the simulated curve with the observed radio flux $S_\nu$. We observed the binary in the orbital plane from apocenter. The time axis is obtained by dividing the distance from the supernova progentior with the velocity of the supernova shell, for which we adopted $v_{sn} = 10^4$\,km/s.}
\label{fig:SN1979C_M15MSunP2000yr_e0_m-5MJ_Snu}
\end{figure}

Figure \ref{fig:SN1979C_M15MSunP2000yr_e0_m-5MJ_Snu} shows the radio lightcurve inferred from the simulation. In this figure we view the binary in the plane of motion from apocenter. From this viewing angle the simulation results seem to agree with the observation, but similar agreement can be achieved for different azimuthal angles.
The red curve in fig.\,\ref{fig:SN1979C_M15MSunP2000yr_e0_m-5MJ_Snu} is calculated using the mean of 100 SPH particles in a cylinder with 3000\,au radius in the direction of the observer. We scaled the distance from the centre of mass by the reciprocal of the velocity, such that each time represents the arrival time of the supernova ejecta to that distance, using a velocity of $v_{sn} =10^4$\,km/s.

We note that the periodic signal in the radio can only be observed in a limited range in time. Before roughly 1000 days, the radio emission is optically thick, in which case the flux is independent of the value of the density. After about 4000 days the ram pressure of the wind becomes comparable to the ambient pressure of the ISM, which breaks our assumption of a freely expanding wind. Therefore, we only expect our model to fit the data between 1000 and 4000 days.

Comparing the observed radio data with the simulations is not trivial because of the confounding factors in the model. These confounding factors include the companion mass, inclination of the viewing angle and the angle along the orbital plane. Changing the two angles introduces rotations in the spiral pattern, which causes a shift in the fluctuations. Changing the companion mass, from about 3\,\MSun\, to $\sim 15$\,\MSun\, causes the density fluctuations in the wind to become more pronounces.
However, the fluctuations in the density variations can be suppressed somewhat by changing the viewing angle. As a consequence, it turns out hard to constrain either the viewing angle or the companion mass.

\begin{figure}
\centering
\includegraphics[width=\columnwidth]{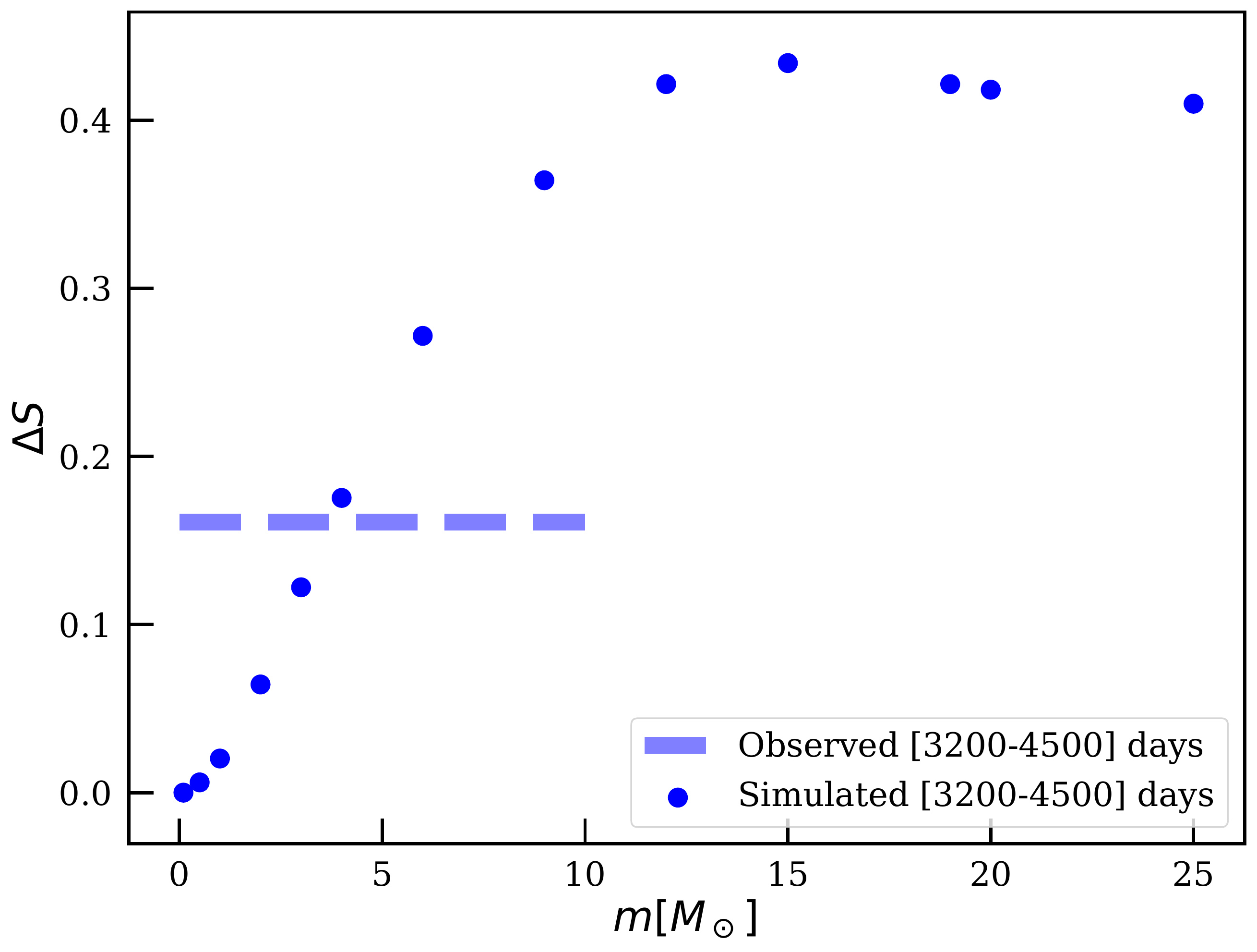}
\caption{The relative difference between the lowest value and the highest values of the density around the second bump at about 4000 days (see Fig.\,\ref{fig:SN1979C_M15MSunP2000yr_e0_m-5MJ_Snu}). The dashed line gives the measured difference in the observations, the bullet points represent the measurements from the simulations for circular orbits and viewed in the plane of motion from the apoapse at an age of 14\,000 yr after the start of the simulation.
}
\label{fig:Mcomp_vs_dSnu}
\end{figure}

To quantify the magnitude of the density contrast in the ripples we present in fig.\,\ref{fig:Mcomp_vs_dSnu} the relative height of the fluctuations $\Delta S = 2 \frac{\rho_{\max} - \rho_{\min}}{\rho_{\max} + \rho_{\min}}$ as a function of companion mass (from 1\,\MSun\, to 25\,\MSun). Here for stars more massive than the primary we adopted a point-mass, rather than an extended stellar object since we expect stars more massive than the supernova progenitor to have collapsed to a black hole at an earlier epoch. 

We further tested the effect of the initial eccentricity of the orbit, varying it from $e=0$ to $e=0.9$.  Up to an eccentricity of $\sim 0.6$, the results did not appreciably change, but higher eccentricities suppress the ripples at larger distance from the star. This is because the stars spend most of the time  near apocenter. In the high-resolution simulations some appreciable deviations from the spiral structure is noticeable when viewed directly from pericenter as respect to other viewing angles, but the effect is much smaller than the fluctuations in the radio observations. We therefore cannot constrain the binary eccentricity in the observations.

\begin{figure}
\centering
\includegraphics[width=\columnwidth]{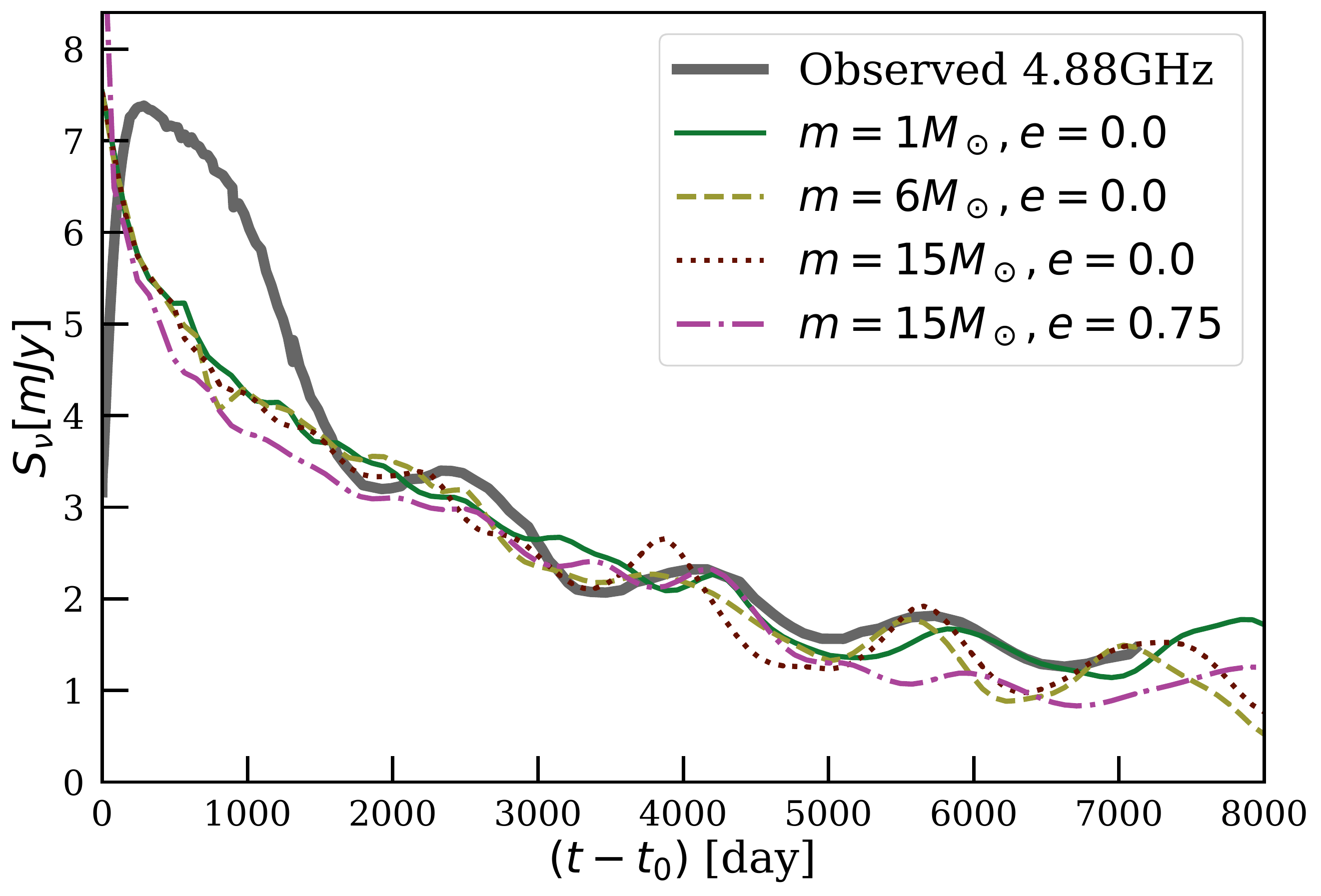}
\caption{
Comparison of four simulations with the same resulting but somewhat different parameters. 
The thick dark-grey curve gives the fit presented in Fig 3 of \citet{Montes2000RadioEvolution} to the radio data of \citet{Weiler1991The1979C}.
As indicated in the top-right corner of the figure, we vary the companion mass of 1\,\MSun, 6\,\MSun\, and 15\,\MSun\, and for circular orbits as well as for one with an eccentricity of $e=0.75$.
All simulated binaries had an orbital period of 2000\,yr, were perfomed at a resolution of $10^{-5}$\,\MSun\, per SPH particle and adopted a viewing angel in the plane of motion and along the binary's long axis (from apocenter).
}

\label{fig:M1and15MSun_for_e0to075}
\end{figure}

In fig.\,\ref{fig:M1and15MSun_for_e0to075} we show the results of a few extra simulations in which we varied the companion mass and the orbital eccentricity. The density variations (and therefore also the fluctuations in radio) are rather pronounced in each of these cases. The intrinsic degeneracy in the problem in the inclination and orbital phase make it hard to constraint the orbital eccentricity or companion masses. The simulation with a shorter orbital period, however, shows more distinct peaks than the observations, and they do not line up nicely as is the case with the 2000yr orbital period.
But this parameter is degenerate in the velocity of the supernova ejecta.

\section{Discussion and conclusions}\label{Sect:discussion}

We propose that the ripples observed in the 4.88GHz and 1.47GHz radio observations in SN1979C result from the interaction between the stellar wind of the progenitor star and the supernova outflow. In that case, the observed ripples originate when the supernova outflow interacts with the slow wind from the progenitor star. Density fluctuations in the wind then cause small variations in the radio flux. The observed variations then relate directly and linearly to density fluctuations in interaction between the progenitor's wind and its the supernova blast wave. We developed this hypothesis by deriving qualitatively the fluctuations in time-scale as well as the amplitude of the variations and associate them to the range in possible binary companions masses and orbital periods. 

We simulated the binary for 14\,000yr before the supernova explosion, taking the binary orbit and the wind produced by the supernova progenitor into account. For this purpose we couple a smoothed particle hydrodynamics solver with a gravitational $N$-body code using the AMUSE software framework. The simulation results agree both with the observations and the theoretical analysis. In addition, the simulations allow us to consider cases where companion mass is comparable to the primary. We cannot constrain the viewing angle, because this parameter is degenerate with the companion mass. The resulting signal is rather insensitive to the orbital eccentricity of the progenitor binary star. The best match with the observations is obtained for a  companion mass between  5\,\MSun\, and about 12\,\MSun\, in a $\sim 2000$\,yr binary with relatively low ($e\aplt 0.8$) eccentricity. 

We conclude that the ripples observed in SN1979C are a natural consequence of a supernova in a binary system. The orbital period, velocity of the stellar wind prior to the supernova and the velocity of the supernova ejecta are directly derivable from the observed ripples in the radio flux. We can only place a lower bound on the mass of the companion star, because of degeneracy with inclination. We argue that a mass ratio of $M/m \apgt 0.3$ is sufficient to produce pronounced ripples in the stellar wind of the supernova progenitor. The viewing angle is hard to constrain. 

The derived binary parameters for the progenitor of SN1979C are rather common for massive stars, and many supernovae are expected to show similar ripples in their radio lightcurves. Those ripples are probably most pronounced between the moment the blast wave becomes optically thin (after about 1000\,days) and the moment the density in the stellar wind becomes comparable to the background density (about $\sim 4000$\,days after the supernova explosion).
The relation between the time between ripples gives a direct measurement of the binary period before the supernova. Unless better observational data becomes available, the amplitude of the fluctuations and the shape of the curves will be insufficient to further  constrain the companion mass or the orbital eccentricity.

\section*{Acknowledgements}

We thank Re'em Sari, Ehud Nakar, Assaf Horesh and Gregg Hallinan for discussions, and Edwin van der Helm for his elaborate work on the AMUSE wind module. We also thank the anonymous referee for helping improve this manuscript.
The calculations were performed on the LGM-II supercomputer at Leiden Observatory (NWO grant \#621.016.701).
SPZ is grateful for the hospitality and support of the Canadian Institute for Theoretical Astrophysics, in particular to Norm Murray. In this work we use the matplotlib \citep{Hunter2007Matplotlib:Environment}, numpy \citep{Oliphant2006ANumPy} and amuse \cite{PortegiesZwart2011AMUSE:Environment} packages.




\bibliographystyle{mnras}
\bibliography{references} 








\bsp	
\label{lastpage}
\end{document}